\documentclass[fleqn,10pt]{wlscirep}
\usepackage{amsthm}
\usepackage{mathtools}
\usepackage{physics}
\usepackage{xcolor}
\usepackage{graphicx}
\usepackage{adjustbox}
\usepackage{placeins}
\usepackage[english]{babel}
\usepackage[utf8]{inputenc}
\usepackage[colorinlistoftodos, color=green!40, prependcaption]{todonotes}
\usepackage[T1]{fontenc}
\usepackage{lipsum}
\usepackage{csquotes}

\usepackage[subrefformat=parens,labelformat=parens,caption=false]{subfig}
\usepackage{collref}

\title{Many-excitation removal of a transmon qubit using a single-junction quantum-circuit refrigerator and a two-tone microwave drive}

\author[1,*]{Wallace Teixeira}
\author[1]{Timm M\"orstedt}
\author[1]{Arto Viitanen}
\author[1]{Heidi Kivijärvi}
\author[1]{Andr\'as Gunyh\'o}
\author[1]{Maaria Tiiri}
\author[1]{Suman Kundu}
\author[1]{Aashish Sah}
\author[1]{Vasilii Vadimov}
\author[1,2]{Mikko M\"ott\"onen}
    
\affil[1]{QCD Labs, QTF Centre of Excellence, Department of Applied Physiscs, Aalto University, P.O. Box 13500, FI-00076 Aalto, Finland}
\affil[2]{VTT Technical Research Centre of Finland Ltd., QTF Center of Excellence, P.O. Box 1000, FI-02044 VTT, Finland}

\affil[*]{wallace.santosteixeira@aalto.fi}

\begin{abstract}
Achieving fast and precise initialization of qubits is a critical requirement for the successful operation of quantum computers. The combination of engineered environments with all-microwave techniques has recently emerged as a promising approach for the reset of superconducting quantum devices. In this work, we experimentally demonstrate the utilization of a single-junction quantum-circuit refrigerator (QCR) for an expeditious removal of several excitations from a transmon qubit. The QCR is indirectly coupled to the transmon through a resonator in the dispersive regime, constituting a carefully engineered environmental spectrum for the transmon. Using single-shot readout, we observe excitation stabilization times down to roughly 500~ns, a 20-fold speedup with QCR and a simultaneous two-tone drive addressing the $e$--$f$ and $f0$--$g1$ transitions of the system. Our results are obtained at a $48$-mK fridge temperature and without postselection, fully capturing the advantage of the protocol for the short-time dynamics and the drive-induced detrimental asymptotic behavior in the presence of relatively hot other baths of the transmon. We validate our results with a detailed Liouvillian model truncated up to the three-excitation subspace, from which we estimate the performance of the protocol in optimized scenarios, such as cold transmon baths and fine-tuned driving frequencies. These results pave the way for optimized reset of quantum-electric devices using engineered environments and for dissipation-engineered state preparation.
\end{abstract}
 
\begin{document}


\global\long\def\ket#1{|#1\rangle}%

\global\long\def\Ket#1{\left|#1\right>}%

\global\long\def\bra#1{\langle#1|}%

\global\long\def\Bra#1{\left<#1\right|}%

\global\long\def\bk#1#2{\langle#1|#2\rangle}%

\global\long\def\BK#1#2{\left\langle #1\middle|#2\right\rangle }%

\global\long\def\kb#1#2{\ket{#1}\!\bra{#2}}%

\global\long\def\KB#1#2{\Ket{#1}\!\Bra{#2}}%

\global\long\def\mel#1#2#3{\bra{#1}#2\ket{#3}}%

\global\long\def\MEL#1#2#3{\Bra{#1}#2\Ket{#3}}%

\global\long\def\n#1{|#1|}%

\global\long\def\N#1{\left|#1\right|}%

\global\long\def\ns#1{|#1|^{2}}%

\global\long\def\NS#1{\left|#1\right|^{2}}%

\global\long\def\nn#1{\lVert#1\rVert}%

\global\long\def\NN#1{\left\lVert #1\right\rVert }%

\global\long\def\nns#1{\lVert#1\rVert^{2}}%

\global\long\def\NNS#1{\left\lVert #1\right\rVert ^{2}}%

\global\long\def\ev#1{\langle#1\rangle}%

\global\long\def\EV#1{\left\langle #1\right\rangle }%

\global\long\def\ha{\hat{a}}%

\global\long\def\hb{\hat{b}}%

\global\long\def\hc{\hat{c}}%

\global\long\def\hd{\hat{d}}%

\global\long\def\he{\hat{e}}%

\global\long\def\hf{\hat{f}}%

\global\long\def\hg{\hat{g}}%

\global\long\def\hh{\hat{h}}%

\global\long\def\hi{\hat{i}}%

\global\long\def\hj{\hat{j}}%

\global\long\def\hk{\hat{k}}%

\global\long\def\hl{\hat{l}}%

\global\long\def\hm{\hat{m}}%

\global\long\def\hn{\hat{n}}%

\global\long\def\ho{\hat{o}}%

\global\long\def\hp{\hat{p}}%

\global\long\def\hq{\hat{q}}%

\global\long\def\hr{\hat{r}}%

\global\long\def\hs{\hat{s}}%

\global\long\def\hu{\hat{u}}%

\global\long\def\hv{\hat{v}}%

\global\long\def\hw{\hat{w}}%

\global\long\def\hx{\hat{x}}%

\global\long\def\hy{\hat{y}}%

\global\long\def\hz{\hat{z}}%

\global\long\def\hA{\hat{A}}%

\global\long\def\hB{\hat{B}}%

\global\long\def\hC{\hat{C}}%

\global\long\def\hD{\hat{D}}%

\global\long\def\hE{\hat{E}}%

\global\long\def\hF{\hat{F}}%

\global\long\def\hG{\hat{G}}%

\global\long\def\hH{\hat{H}}%

\global\long\def\hI{\hat{I}}%

\global\long\def\hJ{\hat{J}}%

\global\long\def\hK{\hat{K}}%

\global\long\def\hL{\hat{L}}%

\global\long\def\hM{\hat{M}}%

\global\long\def\hN{\hat{N}}%

\global\long\def\hO{\hat{O}}%

\global\long\def\hP{\hat{P}}%

\global\long\def\hQ{\hat{Q}}%

\global\long\def\hR{\hat{R}}%

\global\long\def\hS{\hat{S}}%

\global\long\def\hT{\hat{T}}%

\global\long\def\hU{\hat{U}}%

\global\long\def\hV{\hat{V}}%

\global\long\def\hW{\hat{W}}%

\global\long\def\hX{\hat{X}}%

\global\long\def\hY{\hat{Y}}%

\global\long\def\hZ{\hat{Z}}%

\global\long\def\hap{\hat{\alpha}}%

\global\long\def\hbt{\hat{\beta}}%

\global\long\def\hgm{\hat{\gamma}}%

\global\long\def\hGm{\hat{\Gamma}}%

\global\long\def\hdt{\hat{\delta}}%

\global\long\def\hDt{\hat{\Delta}}%

\global\long\def\hep{\hat{\epsilon}}%

\global\long\def\hvep{\hat{\varepsilon}}%

\global\long\def\hzt{\hat{\zeta}}%

\global\long\def\het{\hat{\eta}}%

\global\long\def\hth{\hat{\theta}}%

\global\long\def\hvth{\hat{\vartheta}}%

\global\long\def\hTh{\hat{\Theta}}%

\global\long\def\hio{\hat{\iota}}%

\global\long\def\hkp{\hat{\kappa}}%

\global\long\def\hld{\hat{\lambda}}%

\global\long\def\hLd{\hat{\Lambda}}%

\global\long\def\hmu{\hat{\mu}}%

\global\long\def\hnu{\hat{\nu}}%

\global\long\def\hxi{\hat{\xi}}%

\global\long\def\hXi{\hat{\Xi}}%

\global\long\def\hpi{\hat{\pi}}%

\global\long\def\hPi{\hat{\Pi}}%

\global\long\def\hrh{\hat{\rho}}%

\global\long\def\hvrh{\hat{\varrho}}%

\global\long\def\hsg{\hat{\sigma}}%

\global\long\def\hSg{\hat{\Sigma}}%

\global\long\def\hta{\hat{\tau}}%

\global\long\def\hup{\hat{\upsilon}}%

\global\long\def\hUp{\hat{\Upsilon}}%

\global\long\def\hph{\hat{\phi}}%

\global\long\def\hvph{\hat{\varphi}}%

\global\long\def\hPh{\hat{\Phi}}%

\global\long\def\hch{\hat{\chi}}%

\global\long\def\hps{\hat{\psi}}%

\global\long\def\hPs{\hat{\Psi}}%

\global\long\def\hom{\hat{\omega}}%

\global\long\def\hOm{\hat{\Omega}}%

\global\long\def\hdgg#1{\hat{#1}^{\dagger}}%

\global\long\def\cjg#1{#1^{*}}%

\global\long\def\hsgx{\hat{\sigma}_{x}}%

\global\long\def\hsgy{\hat{\sigma}_{y}}%

\global\long\def\hsgz{\hat{\sigma}_{z}}%

\global\long\def\hsgp{\hat{\sigma}_{+}}%

\global\long\def\hsgm{\hat{\sigma}_{-}}%

\global\long\def\hsgpm{\hat{\sigma}_{\pm}}%

\global\long\def\hsgmp{\hat{\sigma}_{\mp}}%

\global\long\def\dert#1{\frac{d}{dt}#1}%

\global\long\def\dertt#1{\frac{d#1}{dt}}%

\global\long\def\Tr{\text{Tr}}%

\flushbottom
\maketitle
%
%
\thispagestyle{empty}


\section*{Introduction}

The rapid progress in applications~\cite{Kim2023,Mi2021,Arute2020} of superconducting quantum computers brings not only a critical need for improved hardware components, but also optimized techniques for the control of their basic building blocks, the qubits. From the constantly increasing complexity of the quantum processors to the ultimate goal of quantum error correction, the need for fast and accurate initialization of the device to a known eigenstate is a requirement. In the context of noisy intermediate-scale quantum (NISQ) devices, the efficient reuse of qubits~\cite{DeCross2023} emerges as a new paradigm connecting the search for new quantum algorithm designs with optimized reset schemes.

Although unconditional and straightforward, the passive relaxation towards thermal equilibrium is notably inefficient or even insufficient for reset, especially given the demand for increasing qubit lifetimes in advanced operations. To evade the initialization errors arising from unwanted excitation to higher eigenstates and the feedback imposed by measurement-based protocols, parameter tunability plays an important role in reset strategies~\cite{Basilewitsch2021}. Recent reset schemes for superconducting circuits employ this tunability through different ways, such as frequency modulation~\cite{Tuorila2017,Zhou2021}, all-microwave control~\cite{Geerlings2013,Egger2018,Magnard2018}, and engineered environments~\cite{Tan2017,Partanen2018,Basilewitsch2019,Aamir2023,Sevriuk2022}. 

In this work, we experimentally combine the latter two approaches to demonstrate enhanced removal of excitations from a transmon qubit. On one hand, microwave driving allows for excitation transfer from the transmon to a dispersively coupled resonator in a cavity-assisted Raman process~\cite{Zeytinoglu2015}, and on the other hand, the so-called quantum-circuit refrigerator (QCR) coupled to the resonator promotes an overall increase in decay rates through photon-assisted quasiparticle tunneling~\cite{Silveri2017,Hsu2020}, with a minimal effect on qubit lifetimes in its off-state. 

The key component of the QCR is a normal metal-insulator-superconductor (NIS) junction, a structure previously utilized in applications beyond quantum computing, such as thermometry and electronic refrigeration~\cite{Giazotto2006}. Proper control of the bias voltage through this junction results in the incoherent removal or injection of photons into its surroundings, effectively rendering the QCR a controllable environment for its coupled circuitry. Previous experimental studies with the QCR have already shown a speedup of dissipation up to almost four orders of magnitude at the nanosecond scale~\cite{Sevriuk2019,Sevriuk2022}, yielding promising insights into its potential for reset applications. Therefore, integrating the QCR into the hardware of superconducting-based quantum computers may enable its practical utilization in active reset protocols, as we demonstrate in this work.

Our work is motivated by the recent experimental realization of transmon initialization using a QCR and two-tone driving~\cite{Yoshioka2023}, showing a $99.5\%$-reset-fidelity in $180$ ns if no leakage to high-energy states is considered. Here, we expand the approach by showing the excitation removal also from the second excited state of the transmon and by studying the detrimental effects owing to relatively high temperatures. In contrast to Ref.~\cite{Yoshioka2023}, we employ single-shot qubit readout, allowing us to individually capture the contributions arising from the different transmon states.
Moreover, the design of our sample is fundamentally different from that measured in Ref.~\cite{Yoshioka2023}. In addition to the presence of a dedicated readout resonator, our QCR structure is based on a single NIS junction embedded in an auxilliary resonator, allowing for mitigation of charge noise and relief in parameter optimization for the readout and reset compared with the double-junction QCR design~\cite{Vadimov2022}. 

Our experimental results demonstrate not only a moderate speedup in transmon decay dynamics facilitated by the QCR pulse but also a $20$-fold enhancement in short-time excitation decay owing to additional two-tone driving, even in the presence of a relatively hot transmon bath. Importantly, we use a theoretical model that extends the multilevel structure of the transmon--auxiliary-resonator system up to the three-excitation subspace, encompassing the second and third excited states of the transmon. In the case of an extremely cold transmon bath and fine-tuned driving frequencies, this model predicts an asymptotic excitation probability $\sim\!10^{-8}$, even with conservative QCR pulse amplitudes. 

\section*{Results}
\subsection*{General description}

\begin{figure*}
    \subfloat{\label{fig:modela}} 
    \subfloat{\label{fig:modelb}}
    \subfloat{\label{fig:modelc}}
    \subfloat{\label{fig:modeld}}
    \centering
    \includegraphics[width=\linewidth]{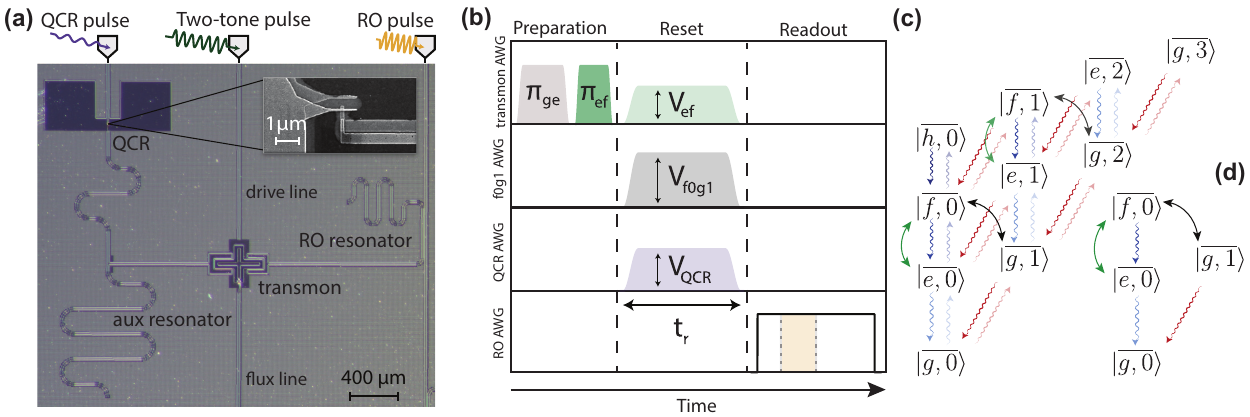}
    \caption{Device, pulse protocol, and energy-level diagram of the system. \textbf{(a)} Micrograph of a sample similar to that utilized in the experiments together with a simplified driving setup. A transmon is coupled to a readout resonator and to an auxiliary resonator which is also coupled to the single-junction quantum-circuit refrigerator (QCR) detailed in the inset. Additional transmon control is provided through dedicated drive and flux lines. \textbf{(b)} Pulse protocol describing the state preparation, the QCR-assisted two-tone reset, and the readout of the system. The transmon is prepared in its first or second excited state through a sequence of $\pi$ pulses. During the reset stage of length $t_{\rm{r}}$, the $e$--$f$ and $f0$--$g1$ transitions of the transmon--auxilary-resonator system are activated with a two-tone drive with adjustable amplitudes $V_{\rm{ef}}$ and $V_{\rm{f0g1}}$, respectively. Simultaneously, the QCR is operated with a net-zero sinusoidal pulse with a fixed frequency and an adjustable amplitude $V_{\rm{QCR}}$. Subsequently, a 2.0-{\textmu}s pulse is applied for the dispersive readout, from which the transmon populations are estimated through a four-component Gaussian mixture model describing the single-shot data in the IQ plane integrated from $0.4$ to $1.0$ {\textmu}s (see the Methods). \textbf{(c, d)} Effective multilevel structure of the transmon--auxiliary-resonator system during the reset stage truncated to the ten \textbf{(c)} and four \textbf{(d)} lowest-energy eigenstates of the system. The green and black bidirectional arrows represent the two-tone-drive-induced coupling between the levels $\ket{\overline{e,n}}\leftrightarrow\ket{\overline{f,n}}$ and $\ket{\overline{f,n}}\leftrightarrow\ket{\overline{g,n+1}}$, respectively. The curly unidirectional arrows indicate the incoherent transitions of the system, which can be potentially modified by the QCR pulse.}
    \label{fig:model}
\end{figure*}

To demonstrate the QCR-expedited excitation removal with the two-tone drive, we carry out experiments with the sample depicted in Fig.~\ref{fig:modela}. Here, a transmon qubit is coupled to the flux and drive lines, and to a quarter-wave readout resonator of frequency $\omega_{\text{RO}}/(2\pi)=7.437$~GHz. To prevent undesirably high decay rates when not biased, the single-junction QCR is indirectly coupled to the transmon through an auxiliary quarter-wave resonator of frequency $\overline{\omega}_{\text{r}}/(2\pi)= 4.671$~GHz. Both resonators are designed to have relatively low quality factors and well-separated frequencies among each other yet within the strong dispersive regime with respect to the transmon frequency. The QCR can be carefully biased with both dc and rf signals to operate in the cooling regime, where the system temperature may be below the normal-metal temperature. This bias point also promotes increased decay rates on demand to the low-temperature bath defined by the QCR.  

The sample is fabricated on a six-inch prime-grade intrinsic-silicon wafer. The superconducting circuit consists of a 200-nm Nb film patterned by dry-etching and the Josephson-junctions are deposited by shadow evaporation using in-situ oxidation to create the tunnel barrier. The NIS junction is fabricated in a similar way, replacing Al with Cu on the top electrode. All measurements are carried out with a dilution refrigerator at the base temperature of $48$~mK and at the flux sweet spot of the transmon. The measured and simulated parameters are provided in Table~\ref{tb:params}. Previous characterization of this sample has also been reported in Ref.~\cite{Viitanen2024}. 
\begin{table}[ht]
  \centering
  \caption{Summary of the relevant experimental and simulation parameters. The superscripts `off' and `on' highlight parameters obtained in the QCR-off and QCR-on states, respectively. The asterisk (*) refers to the parameter values estimated from fitting the numerical model to the experimental data or predicted from theory, see the Methods for details.}
  \label{tb:params}
  \begin{tabular}{ccc}
    \toprule
    \toprule
    {Parameter} & {Symbol} & {Value} \\
    \midrule
    \begin{tabular}{c}
    \underline{transition frequency}\\
    $g$--$e$\\
    $e$--$f$\\
    $f$--$h$\\
    $f0$--$g1$\\
    aux resonator\\
    RO resonator
    \end{tabular} & \begin{tabular}{c}
    {}\\
    $\overline{\omega}_{\text{ge}}/(2\pi)$\\
    $\overline{\omega}_{\text{ef}}/(2\pi)$\\
    $\overline{\omega}_{\text{fh}}/(2\pi)$\\
$\overline{\omega}^{\text{off(on)}}_{\text{f0g1}}/(2\pi)$\\
$\overline{\omega}^{\text{off(on)}}_{\text{r}}/(2\pi)$\\
    $\omega_{\text{RO}}/(2\pi)$
    \end{tabular} & \begin{tabular}{c}
    {}\\
    $4.089$ GHz\\
    $3.816$ GHz\\
    $3.486$ GHz\\
    $3.230\ (3.231)$ GHz\\
    $4.671\ (4.670)$ GHz\\
    $7.437$ GHz\end{tabular}\\
    \midrule
    \begin{tabular}{c}
    \underline{decay time, occupation number}\\
    $e$--$g$\\
    $f$--$e$*\\
    $h$--$f$*\\
    aux resonator*
    \end{tabular} & \begin{tabular}{c}
    {}\\
    $1/\gamma_{\text{eg}}^{\text{off(on)}}$,\ \  $\bar{n}_{\text{eg}}$\\
    $1/\gamma_{\text{fe}}^{\text{off(on)}}$,\ \  $\bar{n}_{\text{fe}}$\\
    $1/\gamma_{\text{hf}}^{\text{off(on)}}$,\ \  $\bar{n}_{\text{hf}}$\\
$1/\kappa^{\text{off(on)}}$,\ \  $\bar{n}_{\text{r}}$
    \end{tabular} & \begin{tabular}{c}
    {}\\
    $6.6\ (4.9)$ {\textmu}s,\ \  $0.20$\\
    $3.3\ (2.5)$ {\textmu}s,\ \  $0.23$\\
    $2.2\ (1.6)$ {\textmu}s,\ \  $0.28$\\
    $221\ (120)$ ns,\ \  $0.15$\end{tabular}\\
    \midrule
    \begin{tabular}{c}
    \underline{dephasing time}\\
    $e$--$g$*\\
    $f$--$e$*\\
    $h$--$f$*
    \end{tabular} & \begin{tabular}{c}
    {}\\
    $1/\gamma_{\text{eg}}^{\phi,\text{off(on)}}$\\
    $1/\gamma_{\text{fe}}^{\phi,\text{off(on)}}$\\
    $1/\gamma_{\text{hf}}^{\phi,\text{off(on)}}$
    \end{tabular} & \begin{tabular}{c}
    {}\\
    $1.7\ (1.2)$ {\textmu}s\\
    $0.8\ (0.6)$ {\textmu}s\\
    $0.6\ (0.4)$ {\textmu}s\end{tabular}\\
    \midrule
    Dynes parameter & $\gamma_{\text{D}}$ & $2.3\times10^{-3}$ \\
    QCR tunneling resistance & $R_{\text{T}}$ & $13.8$ k$\Omega$ \\ 
    \bottomrule
    \bottomrule
  \end{tabular}
\end{table}

We use four arbitrary waveform generators (AWGs) to implement the pulse sequence depicted in Fig.~\ref{fig:modelb}. The transmon is prepared with flat-top $\pi$-pulses, one being addressed to the $g$--$e$ transition, $\overline{\omega}_{\text{ge}}/(2\pi)=4.0885$ GHz, and another subsequently addressed to the $e$--$f$ transition, $\overline{\omega}_{\text{ef}}/(2\pi)=3.8155$ GHz. During the reset stage, two simultaneous microwave drives, with amplitudes $V_{\text{ef}}$ and $V_{\text{f0g1}}$ addressing the $e$--$f$ and $f0$--$g1$ transitions of the transmon--auxiliary-resonator system, respectively, are combined at room temperature, and the signal is applied to the transmon through its drive line. In addition, the QCR is simultaneously biased with a $100$-MHz sine wave pulse of amplitude $V_{\text{QCR}}$ well-below $\Delta/e$, where $e$ is the elementary charge and $\Delta=0.215$~meV is the superconductor gap parameter of aluminium. 
A careful calibration of the pulses is needed for an optimized reset performance as detailed below.

The mechanism underlying the reset protocol is illustrated in Fig.~\ref{fig:modelc}. In the dispersive regime, the transmon and the auxiliary resonator form a hybrid system characterized by the weakly entangled states $\ket{\overline{j,n}}$, where $j=g,e,f,h,...$, and $n=0,1,2,...$ refer to the energy levels of their bare states. The transmon excitations are transferred to the resonator with the help of the two-tone drive, which couples the levels $\ket{\overline{e,n}}\leftrightarrow\ket{\overline{f,n}}$ and $\ket{\overline{f,n}}\leftrightarrow\ket{\overline{g,n+1}}$ with Rabi rates $\Omega_{\rm{ef}}$ and $\Omega_{\rm{f0g1}}\sqrt{n+1}$, respectively. The frequencies  of the two tones are calibrated to be resonant with the transitions $\ket{\overline{e,0}}\leftrightarrow\ket{\overline{f,0}}$ and $\ket{\overline{f,0}}\leftrightarrow\ket{\overline{g,1}}$. 
 Owing to the relatively strong dissipation in the auxiliary resonator, the system rapidly decays to the ground state $\ket{\overline{g,0}}$ in the absence of heating effects. This fast decay is potentially enhanced by the application of the QCR pulse since it may increase the decay rates of the system.

However, at increased temperatures, the effective decoherence channels modified by two-tone driving tend to exhibit a more involved structure due to the multi-level nature of the system, leading to more residual excitations than in the absence of driving as demonstrated below. In this work, we show that the two-tone protocol assisted by the QCR can, nevertheless, provide a substantial speed up of the short-time transmon dynamics, even in the presence of heating and small deviations from the ac-Stark-shifted frequencies. Our results are validated through a Liouvillian model truncated up to the three-excitation subspace with ten states as detailed in the Methods. 

\subsection*{Pulse calibration}

\begin{figure*}
    \subfloat{\label{fig:qcra}} 
    \subfloat{\label{fig:qcrb}}
    \subfloat{\label{fig:qcrc}}
    \centering
    \includegraphics[width=\linewidth]{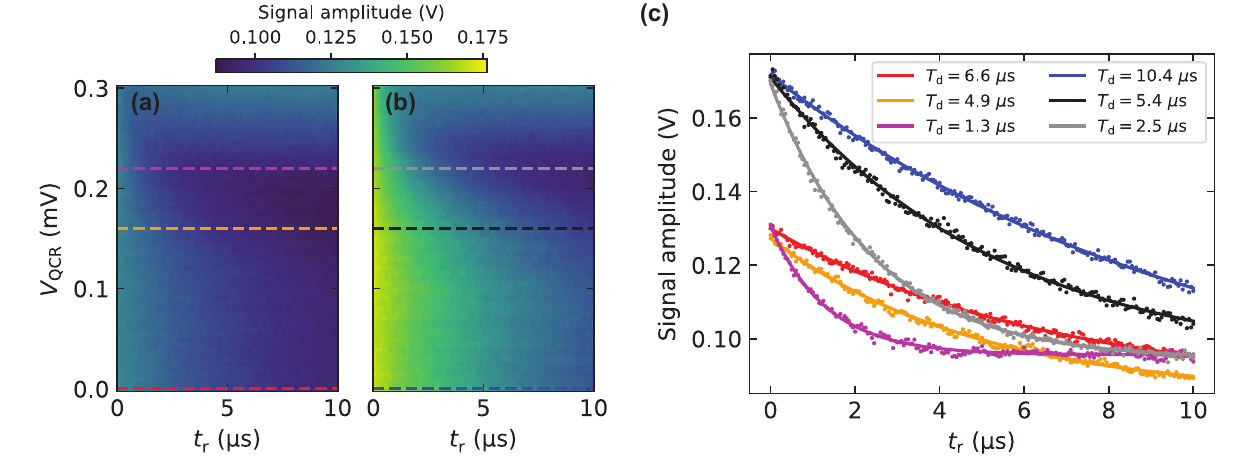}
    \caption{QCR pulse calibration. \textbf{(a, b)} $20,000$-averaged-readout signal amplitude for different QCR pulse amplitudes $V_{\rm{QCR}}$ and reset lengths $t_{\rm{r}}$ in absence of the two-tone drive, $V_{\rm{ef}}=V_{\rm{f0g1}}=0$. The readout frequency is set to $f_{\rm{RO}}=7.4372$ GHz. The panels show data for a state initialized \textbf{(a)} with a single $\pi_{\rm{ge}}$ pulse and \textbf{(b)} with a sequence of $\pi_{\rm{ge}}$ and $\pi_{\rm{ef}}$ pulses. The dashed lines in both plots highlight the amplitudes $V_{\rm{QCR}}=0.0$, $0.16$, and $0.22$ mV. 
    \textbf{(c)} Traces corresponding to the highlighted QCR amplitudes in \textbf{(a, b)} with matching colors. The solid lines represent exponential fits, from which the characteristic stabilization times $T_{\rm{d}}$ are extracted accordingly.}
    \label{fig:qcr}
\end{figure*}

We begin with the calibration of the pulses for the reset protocol by applying a QCR pulse with variable amplitude $V_{\text{QCR}}$ and length $t_{\text{r}}$ in the absence of the two-tone drive as shown in Fig.~\ref{fig:qcr}. The transmon is initialized in its first (Fig.~\ref{fig:qcra}) or second excited state (Fig.~\ref{fig:qcrb}), and the amplitude of the voltage signal transmitted through the readout line is averaged over $20,000$ realizations. For all reset lengths $t_{\text{r}}$, the delay time between the QCR and readout pulses is fixed to $20$~ns. From exponential fits to the data (Fig.~\ref{fig:qcrc}), and for the first excited state, we extract the stabilization times $T_{\rm{d}}=6.6$~{\textmu}s in the QCR-off state, i.e., at $V_{\text{QCR}}=0.0$~mV and $T_{\rm{d}}=4.9$~{\textmu}s in the QCR-on state with amplitude $V_{\text{QCR}}=0.16$~mV. For the second excited state, the stabilization times at the mentioned amplitudes are $T_{\rm{d}}=10.4$~{\textmu}s and $T_{\rm{d}}=5.4$~{\textmu}s, respectively. Roughly, one can think of $T_{\rm{d}}$ as the stabilization timescale of the system after being prepared in a given excited state. Thus, the QCR pulse accelerates the stabilization of the aforementioned prepared states by a factor of $1.35$ and $1.93$, respectively. Note that the averaged signal amplitude is a weighted sum of the quantum-sate probabilities. Consequently, its deviation from the ground-sate value indicates a finite excitation probability of the transmon without distinguishing the individual contributions of the excited states which require single-shot measurements.

The QCR pulse may introduce a trade-off between achieving rapid decay and avoiding heating. As shown in previous work~\cite{Silveri2017,Hsu2020}, for relatively high normal-metal temperatures, the cooling range of the QCR shifts to voltages below $\Delta/e$, where the decay rates are not maximal. From Figs.~\ref{fig:qcra} and~\ref{fig:qcrb}, we observe that the region between $0.16\ \rm{mV}\lesssim V_{\rm{QCR}}\lesssim 0.25\ \rm{mV}$ yields fast decay, but the equilibrium signal amplitude also changes with the pulse. This range is roughly at the top edge of the thermal activation region of the QCR, where thermal excitation is the dominant mechanism of electron tunneling in the NIS junction, giving rise to increasing QCR temperatures. Consequently, although faster stabilization times are obtained at the vicinities of the superconductor gap voltage, e.g., $V_{\text{QCR}}=0.22$ mV, the system tends to stabilize to a different state from that in the QCR-off-state scenario, as suggested by the saturation of the signal amplitude at high values. Therefore, to avoid additional excitation of the transmon, we conservatively set the QCR-on-state amplitude to $V_{\text{QCR}}=0.16$ mV in the subsequent measurements. In the future, the issues of heated stationary states can be relieved by improved thermalization of the normal-metal and of the superconductor of the QCR. We highlight that fast on-demand generation of thermal states can be achieved~\cite{Moerstedt2024} if the QCR operates with voltages beyond $\Delta/e$, however, this is out of the scope of the present work.

\begin{figure*}
    \subfloat{\label{fig:twotonea}} 
    \subfloat{\label{fig:twotoneb}}
    \subfloat{\label{fig:twotonec}}
    \subfloat{\label{fig:twotoned}}
    \subfloat{\label{fig:twotonee}} 
    \subfloat{\label{fig:twotonef}}
    \subfloat{\label{fig:twotoneg}}
    \centering
    \includegraphics[width=\linewidth]{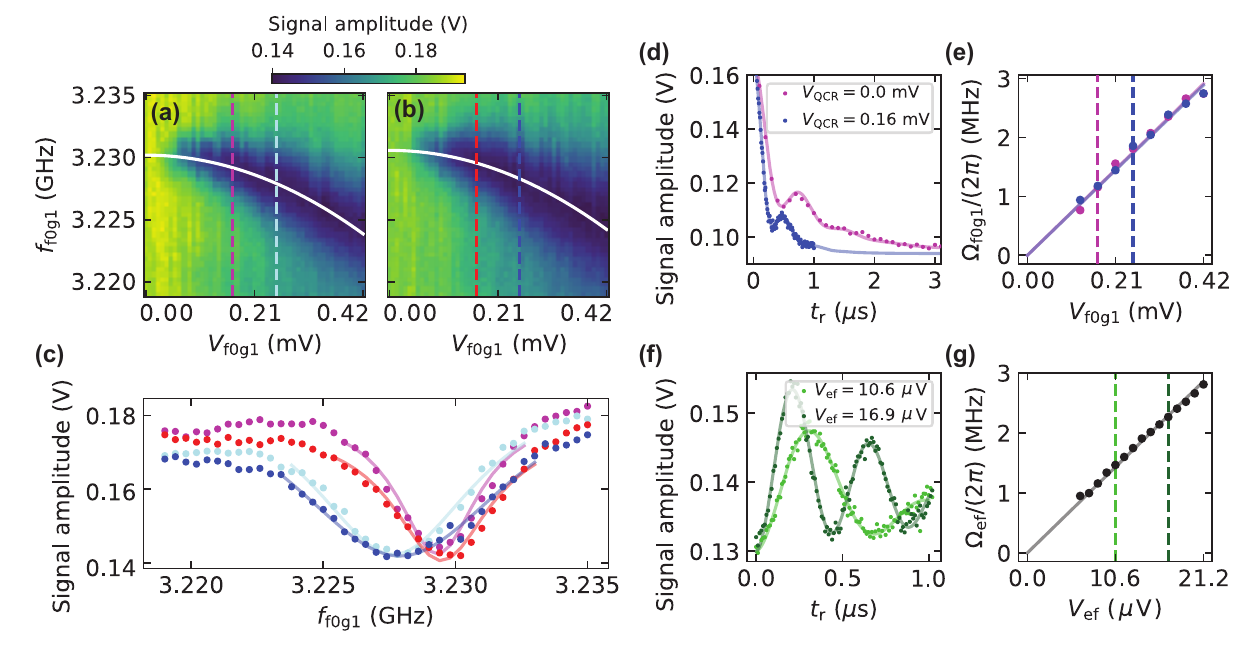}
    \caption{Two-tone drive and Rabi angular frequency calibration. \textbf{(a, b)} Averaged-readout signal amplitude after a 500-ns drive of amplitude $V_{\rm{f0g1}}$ and frequency $f_{\rm{f0g1}}$ in absence of the $e$--$f$ drive, $V_{\rm{ef}}=0$. The system is initialized with a sequence of $\pi_{\rm{ge}}$ and $\pi_{\rm{ef}}$ pulses. The dips in the data indicate the Stark-shifted frequency of the $f0$--$g1$ transition, which is fitted with a quadratic function (solid white curve). The panels show data for QCR amplitudes \textbf{(a)} $V_{\rm{QCR}}=0.0$ and \textbf{(b)} $0.16$ mV. The dashed lines highlight the amplitudes $V_{\rm{f0g1}}=0.17$ and $0.25$ mV.
    \textbf{(c)} Traces corresponding to the highlighted values of $V_{\rm{f0g1}}$ in \textbf{(a, b)}, with the corresponding color codes and asymmetric Lorentzian fits (solid lines). \textbf{(d)} Temporal dependence of the signal amplitude in the presence of the $f0$--$g1$ drive with $V_{\rm{f0g1}}=0.17$ (magenta) and $V_{\rm{f0g1}}=0.25$ mV (blue) for the transmon initialized in the second excited state. The corresponding drive frequencies are chosen to be near the Stark-shifted frequencies extracted from \textbf{(a, b)}. \textbf{(e)} Dependence of the Rabi angular frequency $\Omega_{\rm{f0g1}}$ on the drive amplitude $V_{\rm{f0g1}}$ for the QCR off (magenta) and on (blue). The dashed lines highlight the values of $V_{\rm{f0g1}}$ used in \textbf{(d)}. \textbf{(f)} Temporal variation of the signal amplitude in the presence of the $e$--$f$ pulse and in the absence of the $f0$--$g1$ and QCR pulses, $V_{\rm{f0g1}}=V_{\rm{QCR}}=0$. The transmon is initialized in the first excited state. \textbf{(g)} Dependence of the Rabi angular frequency $\Omega_{\rm{ef}}$ on the drive amplitude $V_{\rm{ef}}$. The dashed lines highlight the values of $V_{\rm{ef}}$ used in \textbf{(f)}. The time-dependent fitting functions of \textbf{(d, f)} are obtained from the theoretical model described in the Methods.}
    \label{fig:twotone}
\end{figure*}

We proceed with the calibration of the two-tone drive as shown in Fig.~\ref{fig:twotone}. First, we initialize the system with a sequence of $\pi_{\text{ge}}$ and $\pi_{\text{ef}}$ pulses. Subsequently, we apply a $500$-ns drive of amplitude $V_{\rm{f0g1}}$ at frequency $f_{\rm{f0g1}}$, and read out the averaged signal amplitude. The drive frequency is swept around the transition frequency $\overline{\omega}_{\text{f0g1}}/(2\pi)$ between the levels $\ket{\overline{f,0}}\leftrightarrow\ket{\overline{g,1}}$.
This procedure is repeated in the QCR-off state (Fig.~\ref{fig:twotonea}) and in the QCR-on state (Fig.~\ref{fig:twotoneb}), both without the $e$--$f$ drive. 

For relatively high amplitudes, we observe that the $f0$--$g1$ drive induces a clear ac-Stark shift of the resonance frequency in both QCR-on and QCR-off states, $\overline{\omega}_{\text{f0g1}}^{\text{on/off}}/(2\pi)$. In both Figs.~\ref{fig:twotonea} and~\ref{fig:twotoneb}, we select the lowest value of the signal amplitude for each $V_{\rm{f0g1}}$ and fit the data to a quadratic function, obtaining the base frequencies $\overline{\omega}_{\text{f0g1}}^{\text{off}}/(2\pi)=3.2302$ GHz, $\overline{\omega}_{\text{f0g1}}^{\text{on}}/(2\pi)=3.2306$ GHz, and negative ac-Stark shifts up to $6.4$ MHz.

The simultaneous QCR pulse (Fig.~\ref{fig:twotoneb}) effectively increases the $f0g1$ resonance frequency up to $0.4$~MHz as a consequence of a negative Lamb shift of the auxiliary resonator frequency~\cite{Viitanen2021}. Additionally, an overall broadening of the spectral line is observed in the QCR-on state. Since the QCR operates with a relatively low voltage bias, we attribute this effect to the increase in the decay rates of the system. Both Lamb shift and line broadening for selected drive amplitudes are shown in Fig.~\ref{fig:twotonec}.

Next, we calibrate the Rabi frequencies of the two-tone drive through time-domain measurements. To extract their dependence on the corresponding pulse amplitudes, we use a Liouvillian model truncated to the fourth level of the system, as detailed in the Methods. 
First, we set $V_{\text{ef}}=0$ and study the decay of the second excited state of the transmon in the presence of $f0$--$g1$ and QCR pulses, as shown in Fig.~\ref{fig:twotoned}.
In general, we observe damped oscillations indicating the rapid excitation decay from the auxiliary resonator. Figure~\ref{fig:twotonee} reveals a good linear dependence between $\Omega_{\text{f0g1}}/(2\pi)$ and $V_{\text{f0g1}}$, with the slope barely affected by the QCR pulse.
Although the $f0$--$g1$ drive frequencies are chosen according to the quadratic fits in Figs.~\ref{fig:twotonea} and~\ref{fig:twotoneb}, the oscillation pattern in Fig.~\ref{fig:twotoned} is also affected by coarse frequency adjustments resulting from the low resolution in the $f0g1$ frequency sweep.
The fitting from the Liouvillian model also enables the estimation of average auxiliary resonator decay rates in the QCR-off and QCR-on states, $1/\kappa^{\text{off}}=221$ ns and $1/\kappa^{\text{on}}=120$ ns, respectively. Importantly, the estimation of these decay rates from time-resolved experiments appeared more convenient and reliable for modeling the reset than usual spectroscopy since the latter requires an additional calibration of the probe power.

A similar calibration is carried out for the $e$--$f$ drive, as shown in Figs.~\ref{fig:twotonef} and~\ref{fig:twotoneg}. The system is initialized in the first excited state, after which damped Rabi oscillations are observed. Here, we set $V_{\text{f0g1}}=V_{\text{QCR}}=0$, such that the oscillations are only attributed to the weak $e$--$f$ drive (Fig~\ref{fig:twotonef}). We observe that also $\Omega_{\rm{ef}}/(2\pi)$ exhibits a linear dependence with $V_{\rm{ef}}$ (Fig~\ref{fig:twotoneg}). Due to the single-photon nature of the $e$--$f$ transition, it requires a much lower driving amplitude for a given Rabi frequency compared to the $f0$--$g1$ transition which involves two-photon processes. Although the $e$--$f$ drive contributes to negligible ac-Stark shifts, its frequency needs to be adjusted in the presence of the $f0$--$g1$ drive.

\subsection*{Excitation dynamics}

After the pulse calibration detailed above, we carry out single-shot readout of the transmon with the pulse configurations described in Table~\ref{tb:configs}. In the configurations $\textbf{A}$ and $\textbf{B}$, the QCR is off, and configurations $\textbf{C}$ and $\textbf{D}$ comprise the QCR-on state. Following the analysis in Ref.~\cite{Magnard2018} for non-Hermitian Hamiltonians, the two-tone voltage amplitudes are chosen as such to produce Rabi angular frequencies in the region where the maximum decay rate of the slowest eigenmode is attained. This requires a strong $f0$--$g1$ drive such that $\Omega_{\rm{f0g1}}\geq2\sqrt{2/27}\kappa$. The additional factor of $2$ arises from the rescaling by $1/2$ of the Rabi amplitudes in our model compared to that in Ref.~\cite{Magnard2018}. 

\begin{figure*}
    \subfloat{\label{fig:sspexca}}
    \subfloat{\label{fig:sspexcb}}
    \subfloat{\label{fig:sspexcc}}
    \subfloat{\label{fig:sspexcd}}
    \subfloat{\label{fig:sspexce}}
    \subfloat{\label{fig:sspexcf}}
    \subfloat{\label{fig:sspexcg}}
    \subfloat{\label{fig:sspexch}}
    \subfloat{\label{fig:sspexci}}
    \subfloat{\label{fig:sspexcj}}
    \subfloat{\label{fig:sspexck}}
    \subfloat{\label{fig:sspexcl}}
    \centering
    \includegraphics[width=\linewidth]{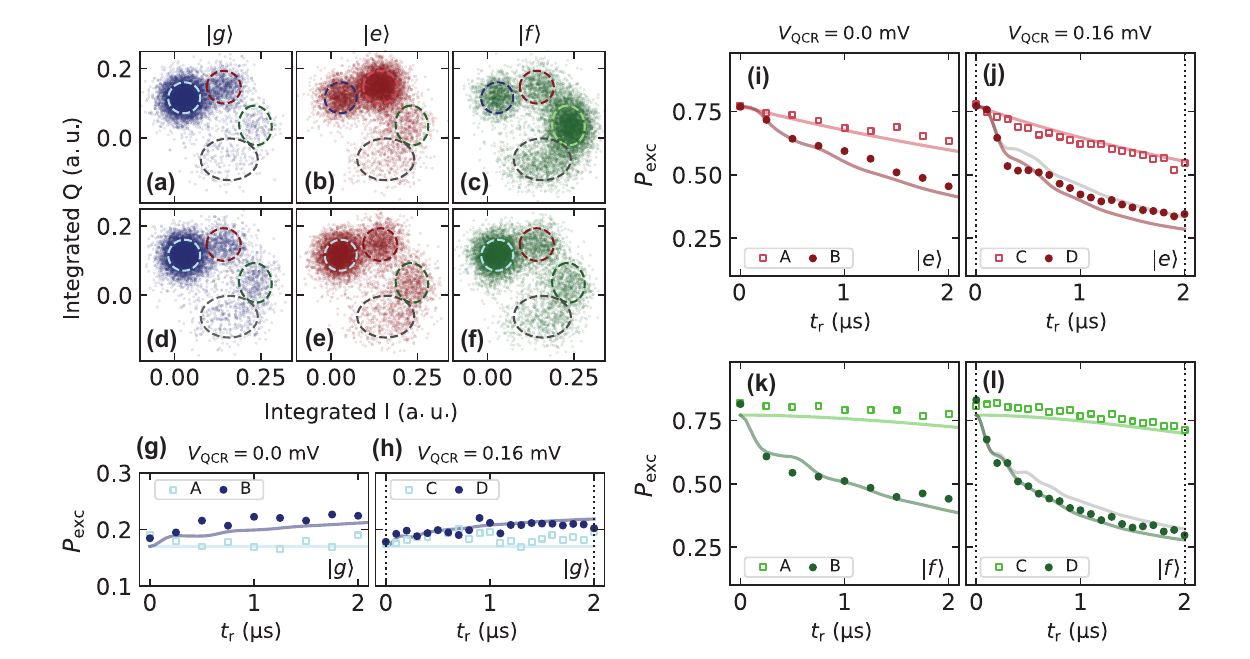}
    \caption{Single-shot data and excitation dynamics. \textbf{(a--f)} Example distributions of $8,000$ single-shot measurements acquired after the reset times \textbf{(a--c)} $t_{\rm{r}}=0$ and \textbf{(d--f)} $t_{\rm{r}}=2$ {\textmu}s for the different prepared states as indicated and with the reset pulse amplitudes in configuration \textbf{D} defined in  Table~\ref{tb:configs}. The labels $\ket{g}$, $\ket{e}$, and $\ket{f}$ refer to the qubit states initialized by starting from the thermal equilibrium, with an additional $\pi_{\rm{ge}}$ pulse, and with a sequence of $\pi_{\rm{ge}}$ and $\pi_{\rm{ef}}$ pulses, respectively. The dashed-line ellipses are obtained from the four-component Gaussian mixture model calibrated at $t_{\rm{r}}=0$. \textbf{(g--l)} Excitation probability $P_{\rm{exc}}$ as a function of the reset time $t_{\rm{r}}$ for different prepared states as indicated. The pulse amplitudes in \textbf{(g, i, k)} are chosen as in configurations \textbf{A} or \textbf{B} and in  \textbf{(h, j, l)} as in \textbf{C} or \textbf{D} of Table~\ref{tb:configs}. Solid curves are obtained from the master equation simulation of the Liouvillian $\mathcal{L}$ truncated at the tenth level with corrections for the Stark shifts of the $\ket{\overline{e,0}}$, $\ket{\overline{g,1}}$, and $\ket{\overline{f,0}}$ levels. Remaining parameters used in the simulations are given in Table~\ref{tb:params}. The grey curves in \textbf{(j, l)} are from the simulation truncated to the tenth level with drives as in configuration \textbf{D}, but with the estimated value of $\kappa$ in the QCR-off state. Dotted vertical lines in \textbf{(h, j, l)} indicate the values of $t_{\textrm{r}}$ in \textbf{(a--f)}.}
    \label{fig:sspexc}
\end{figure*}

Figures~\ref{fig:sspexca}--\ref{fig:sspexcf} show examples of $8,000$ single-shot data points each in the in-phase--quadrature-phase (IQ) plane for configuration $\textbf{D}$. The labels $\ket{g}$, $\ket{e}$, and $\ket{f}$ indicate the ideal initial states, which are prepared or not with the corresponding sequence of $\pi$ pulses. In general, we observe four prominent point clouds associated to the lowest states of the transmon. While driving errors and measurement-induced transitions are common sources of leakage~\cite{Sank2016}, we attribute the emergence of more than two point clouds primarily to the relatively hot transmon environment. These point clouds persist even in the thermal equilibrium state (Fig.~\ref{fig:sspexca}). 
\begin{table}[ht]
  \centering
  \caption{Voltage amplitudes of pulses defined in Fig.~\ref{fig:model} for the reset configurations $\textbf{A}$--$\textbf{D}$ investigated through single-shot measurements. These correspond to the values estimated at the input ports of the sample.}
  \label{tb:configs}
  \begin{tabular}{ccccc}
    \toprule
    \toprule
    {Pulse amplitude ({\textmu}V)} & \textbf{A} & \textbf{B} & \textbf{C} & \textbf{D} \\
    \midrule
    $V_{\text{QCR}}$ & 0.0 & 0.0 & 160.0 & 160.0 \\
    $V_{\text{f0g1}}$ & 0.0 & 169.1 & 0.0 & 253.6 \\
    $V_{\text{ef}}$ & 0.0 & 10.6 & 0.0 & 16.9 \\
    \bottomrule
    \bottomrule
  \end{tabular}
\end{table}

For the reset length $t_{\text{r}}=2.0$ {\textmu}s (Figs.~\ref{fig:sspexcd}--\ref{fig:sspexcf}), the points cluster predominantly in the leftmost ground-state cloud, indicating the excitation removal assisted by the QCR and two-tone drive. We highlight that the single-shot data corresponds to the raw integrated timetraces at an optimized readout frequency, without the use of weighting functions or postselection. Since the readout resonator couples directly to the transmon, the auxiliary resonator states are indistinguishable in the single-shot measurements.

To further investigate the performance of the reset protocol, we study the dynamics of the excitation probability,
\begin{align}
    P_{\text{exc}} = p_{\text{e}} + p_{\text{f}} + p_{\text{h}} = 1- p_{\text{g}},\label{eq:Pexc}
\end{align}
where $p_{j}$ is the probability of the transmon state $\ket{j}$. Experimentally, $p_{j}$ is quantified from the single-shot measurements by a normalized number of data points inside the corresponding ellipse, which is determined using a four-component Gaussian mixture model, see the Methods. Theoretically, we define the transmon populations as
\begin{align}
    p_{j}^{\text{theory}} = \sum_n\mel{\overline{j,n}}{\hrh}{\overline{j,n}},\label{eq:pj_th}
\end{align}
where $\hrh$ is the interaction picture density operator of the transmon--auxiliary-resonator system, and the interaction picture is defined by the multilevel Jaynes-Cummings Hamiltonian. The dynamics of $\hrh$ follows from a Lindblad master equation $\textrm{d}\hrh/\textrm{d}t=\mathcal{L}(\hrh)$, where $\mathcal{L}$ is the Liouvillian superoperator describing the relevant processes during the reset stage, see the Methods.

Figures~\ref{fig:sspexcg}--\ref{fig:sspexcl} show the short-time dynamics of $P_{\text{exc}}$ in the driving configurations of Table~\ref{tb:configs} for different initial states. Starting from the thermal equilibrium state in Figs.~\ref{fig:sspexcg} and~\ref{fig:sspexch}, we observe a weak effect of the QCR pulse in the absence of two-tone driving in configuration \textbf{C}, as $P_{\text{exc}}$ exhibits tiny oscillations near the equilibrium value without the QCR pulse, given in configuration \textbf{A}. From the individual contributions $p_{j}$, the thermal equilibrium temperature of the transmon is determined as $T=110$ mK through fitting $p_{j}$ to the closest Boltzmann distribution, and the average thermal occupation number at the relevant frequencies of the system given in Table~\ref{tb:params}. 
For configurations \textbf{B} and \textbf{D}, the two-tone driving combined with the high transmon temperature induce additional excitations indicated by higher $P_{\text{exc}}$ than that without the drive.

In Figs.~\ref{fig:sspexci}--\ref{fig:sspexcl} where the system is prepared with $\pi$ pulses, we note a moderate speedup of the decay by the QCR pulse alone, as also suggested in Fig.~\ref{fig:qcr}. Despite the coarse frequency tuning and other potential driving errors, the benefits of the two-tone pulse for the short-time removal of $P_{\text{exc}}$ are clearly visible for the considered initial states. In particular, we obtain in Fig.~\ref{fig:sspexcl} a rough stabilization time of $T_{\text{d}}\sim500$ ns for configuration \textbf{D} if the system is initialized in the second excited state. This corresponds to an over 20-fold speedup in comparison to the corresponding intrinsic dynamics.

Figures~\ref{fig:sspexcg}--\ref{fig:sspexcl} exhibit a very good agreement between the experimental data and the master equation simulations. In the absence of two-tone driving, the simulations very accurately reproduce the expected thermal-equilibrium and excitation-decay profiles. Furthermore, our model, which takes into account additional corrections to the energy levels, captures the main dynamics in the configurations with two-tone driving. With the two-tone Rabi frequencies $[\Omega_{\text{f0g1}}/(2\pi),\Omega_{\text{ef}}/(2\pi)]=(1.16,1.43)$~MHz in configuration \textbf{B} and $[\Omega_{\text{f0g1}}/(2\pi),\Omega_{\text{ef}}/(2\pi)]= (1.73,2.29)$~MHz in configuration \textbf{D}, we expect the reset to operate in the underdamped regime in both cases, with reset rates still maximized for these choices~\cite{Magnard2018}. In addition to the state preparation and measurement (SPAM) errors, we attribute the minor discrepancies between the simulations and the experiments to possibly imperfect estimation of the decay rates of the auxiliary resonator and the higher energy levels of the transmon.

The accelerated decay of $P_{\text{exc}}$ produced by the two-tone drive can be connected to the eigenspectrum of the Liouvillian $\mathcal{L}$, denoted by $\{\lambda_j\}$. The spectrum and the corresponding eigenoperators trivially yield the dynamics and the steady state $\hrh_{\rm{ss}}$. Specifically, the presence of at least one null eigenvalue, $\lambda_j=0$, implies the existence of a steady state. The negative real parts of the remaining eigenvalues determine the exponential decay of the system since they are associated to the rates of the corresponding incoherent processes~\cite{Carollo2021,Zhou2023}, namely, $\Lambda_j=|\textrm{Re}(\lambda_j)|$. In general, the stabilization timescale is set by the minimal non-zero rate $\Lambda_j$. In practice, however, the influence of this term in the dynamics depends on the initial state. 

\begin{figure*}
    \subfloat{\label{fig:Lpexcssa}}
    \subfloat{\label{fig:Lpexcssb}} 
    \subfloat{\label{fig:Lpexcssc}}
    \subfloat{\label{fig:Lpexcssd}} 
    \centering
    \includegraphics[width=\linewidth]{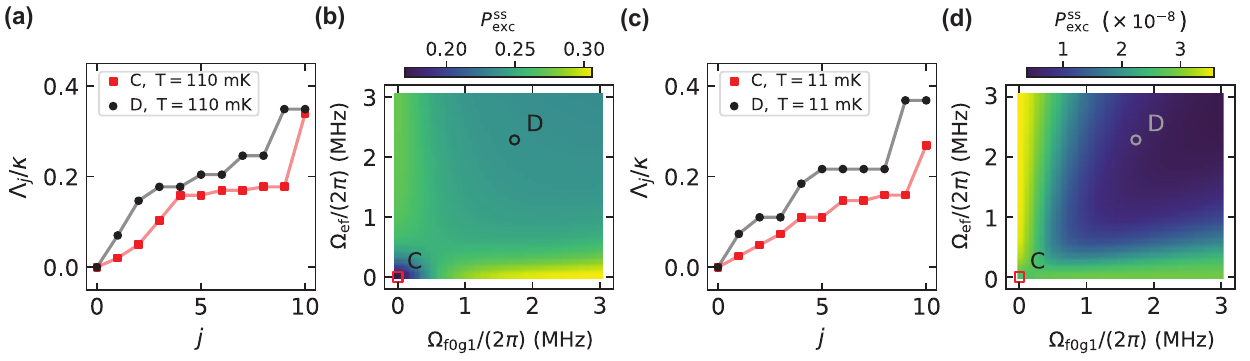}
    \caption{Decay rates and asymptotic excitation probability of the transmon--auxiliary-resonator system. \textbf{(a)} Ten lowest incoherent rates $\Lambda_j$ given by the Liouvillian $\mathcal{L}$ truncated at the tenth level with drive amplitudes in configurations \textbf{C} and \textbf{D} as indicated. The auxiliary-resonator decay rate is denoted by $\kappa$. \textbf{(b)} Asymptotic, long-time excitation probability $P_{\rm{exc}}^{\rm{ss}}$ as a function of the Rabi angular frequencies $\Omega_{\rm{f0g1}}$ and $\Omega_{\rm{ef}}$ for the model truncated at the tenth level. This scenario comprises a relatively hot transmon bath and coarse frequency adjustment as in the modeling of experimental data in Fig.~\ref{fig:sspexc}. 
    \textbf{(c, d)} As \textbf{(a, b)} except for an optimized scenario comprising a cold transmon bath and fine-tuned driving frequencies. In all plots, decay and dephasing rates are chosen as in Fig.~\ref{fig:sspexc}.}
    \label{fig:Lpexcss}
\end{figure*}

Figure~\ref{fig:Lpexcssa} compares the smallest rates $\Lambda_j$ obtained in configurations \textbf{C} and \textbf{D}. An increase in these rates for the latter configuration is observed. Owing to the well-separated timescales set by the different decay rates of the transmon and the auxiliary resonator, we expect the smallest non-zero rates $\Lambda_j$ to be the most relevant in the reset protocol. From our model, a normalized error function $\Delta P_{\rm{exc}}(t_{\rm{r}})\equiv|P_{\rm{exc}}(t_{\rm{r}})-P_{\rm{exc}}^{\rm{ss}}|/|P_{\rm{exc}}(0)-P_{\rm{exc}}^{\rm{ss}}|$ below $10^{-3}$ is reached with $t_{\rm{r}}\gtrsim38.8$ \textmu s in configuration \textbf{C} and for $t_{\rm{r}}\gtrsim7.8$ \textmu s in configuration \textbf{D} for the transmon prepared in the first-excited state. Here, $P_{\text{exc}}^{\text{ss}}$ denotes the asymptotic excitation probability, which is obtained by solving the vectorized equation $\mathcal{L}\hrh_{\rm{ss}}=0$. 

As already suggested in Figs.~\ref{fig:sspexcg} and~\ref{fig:sspexch}, two-tone driving tends to increase $P_{\text{exc}}^{\text{ss}}$ owing to the relatively hot bath, which we also theoretically confirm in Fig.~\ref{fig:Lpexcssb} for a broad range of Rabi frequencies. This effect in our model is caused by the competition between coherent driving and dissipation. For sufficiently strong driving, the steady state of the combined transmon-auxiliary resonator system presents non-vanishing coherences. This compromise can be significantly reduced in the presence of a cold bath and with fine-tuning of the driving frequencies, as illustrated in Figs.~\ref{fig:Lpexcssc} and~\ref{fig:Lpexcssd}. In such cases, two-tone driving can provide substantial speedup in excitation removal with asymptotic values $P_{\text{exc}}^{\text{ss}}\sim\!10^{-8}$.

\section*{Discussion}

We have experimentally demonstrated the many-excitation removal from a transmon qubit assisted by a quantum-circuit refrigerator in combination with a two-tone microwave drive. Following a careful pulse calibration, we observed the short-time dynamics of the transmon populations through single-shot measurements of the different transmon states in various driving configurations and initial states. To the best of our understanding, this constitutes the first experimental realization of a two-tone transmon reset protocol using the QCR with only one junction or single-shot readout.

We observed that the relatively high transmon temperature of $T=110$ mK is barely affected by the QCR-on state with conservative pulse amplitudes, in agreement with previous theory on photon-assisted quasiparticle tunneling. We attribute such a high transmon temperature to the thermalization induced by the different components in the cryogenic environment, along with potential quasiparticle generation in the superconducting circuit~\cite{Houzet2019}. We also observed short coherence times even in the QCR-off state, which we may attribute to fluctuations in the transmon frequencies. Consequently, this posed challenges for studying the QCR effects on the qubit coherences, which can be investigated in future experiments, for example, through the thermal photon-induced qubit dephasing rate or quantum tomography. Importantly, the QCR-assisted cooling of the auxiliary resonator has recently been demonstrated in photon-number-resolving experiments carried out with the same sample used in our work~\cite{Viitanen2024}. These experiments complement our study, providing a clear indication that the QCR enables cooling of the circuit on demand.

Nevertheless, in our studies, the QCR was still able to roughly halve the stabilization time of the system when initialized in the second excited state of the transmon. The stabilization time was further reduced by a factor of $\sim10$ with additional two-tone driving, yet some heating, provoking a decrease in the ground state population was observed and theoretically explained by a simple Liouvillian model. We attribute this effect to the involved multi-level structure of the system showcasing the nontrivial interplay between driving and thermal environment. In principle, a refinement in the reset theory could be achieved, for example, through more sophisticated models taking into account multi-photon decay channels or suitably optimized to treat time-dependent Hamiltonians and dissipative environments~\cite{Teixeira2022}.

In the near future, the reset protocol can be optimized in different ways. 
In addition to achieving colder cryogenic environments through proper electromagnetic shielding and reduced electrical noise, improvements in the QCR design can also play an important role. For example, the inclusion of quasiparticle barriers and copper baths connected to the normal-metal island may pave the way for the realization of low-noise circuits, particularly in the QCR-off state. The resulting lower normal-metal temperature likely allows for a cooling operation with higher voltages close to $\Delta/e$, thereby promoting increased decay rates.
Furthermore, the decay rates can be significantly increased upon the fabrication of QCRs targeting smaller tunneling resistances and smaller Dynes parameters by reducing the cross section of the NIS junction or modifying oxidation parameters. These advancements, aligned with its circuitry simplicity, render the single-junction QCR a potential candidate for an efficient and highly tunable cooling device in future experiments.

In the scenarios above, the optimization of two-tone Rabi frequencies may allow the slowest eigenmode of the system to theoretically decay at the maximum rate $\kappa/3$~\cite{Magnard2018}, or in terms of the Liouvillian model, $\Lambda_1=\kappa/6$, with $\kappa$ being orders of magnitude higher than its value in the QCR-off state. Note that these bounds are derived in the minimum subspace that encompasses the second-excited state of a non-decaying transmon. Consequently, the slowest eigenmode is naturally expected to decay at a higher rate in the presence of additional incoherent processes. More accurate descriptions of the reset dynamics with a minimal dependence on the initial state may incorporate the influence of additional eigenmodes. Therefore, our results pave the way for a detailed understanding of the dissipative dynamics of superconducting qubits with engineered decay channels, potentially allowing the exploration of new reset regimes in the future. 

The combined effects of the QCR and two-tone driving on multi-qubit systems are yet to be demonstrated. Nonetheless, the minimal setup for the studied hybrid reset scheme only requires the integration of QCRs to the superconducting resonators dispersively coupled to the superconducting qubits, in addition to their corresponding drive lines. Therefore, this reset strategy just employs the state-of-art resources available on NISQ devices with a minimal amount of excess control lines. 

\section*{Methods}


\subsection*{Theoretical model}
We model the reset of the transmon--auxiliary-resonator system using the multilevel Jaynes--Cummings (JC) Hamiltonian $\hH_0$, the ladder structure of which is depicted in Figs.~\ref{fig:modelc} and~\ref{fig:modeld}. The spectral decomposition of $\hH_0$ is expressed as $\hH_0=\hbar\sum_{j,n}\overline{\omega}_{{j},n}\ket{\overline{j,n}}\bra{\overline{j,n}}$, where $\hbar \overline{\omega}_{{j},n}$ are the eigenenergies of the system in the absence of driving and $\ket{\overline{j,n}}$ are the corresponding dressed eigenstates. Below, we denote the bare states of the system as $\{\ket{j}\otimes\ket{n}=\ket{j,n}\}$, where $j=\text{g},\text{e},\text{f},\text{h},\dots$, and $n=0,1,2,3\dots$. Owing to the ladder structure of the multilevel JC Hamiltonian, $\ket{j,n}$ can be expressed as a linear combination of the dressed states $\{\ket{\overline{j',n'}}\}$ which involve an equal total number of excitations. 

The reset protocol consists of a $100$-MHz QCR pulse in addition to two simultaneous microwave pulses addressed to the $\ket{\overline{e,0}}\leftrightarrow\ket{\overline{f,0}}$ and $\ket{\overline{g,1}}\leftrightarrow\ket{\overline{f,0}}$ transitions. In the interaction picture with respect to $\hH_0$, we model the open dynamics of the system density operator, $\hrh$, through the Lindblad master equation~\cite{Breuer2007}
\begin{align}
    \frac{\text{d}\hrh}{\text{d}t} = \mathcal{L}_{\rm{c}}(\hrh) + \mathcal{L}_{\downarrow}(\hrh)+\mathcal{L}_{\uparrow}(\hrh)+\mathcal{L}_{\phi}(\hrh),\label{eq:LME0}
\end{align}
where $\mathcal{L}_{\rm{c}}(\hrh)=-\text{i}[\hH',\hrh]/\hbar$ describes the coherent dynamics, with $\hH'=\hH+\hH_{\delta}$ being the effective two-tone-driven system Hamiltonian, and the last three terms characterize the thermal emission, absorption, and pure dephasing, respectively.

Next, we discuss the processes described by Eq.~\eqref{eq:LME0}. First, we assume that the drive operator is proportional to the annihilation operator of the transmon,
\begin{align}
    \hb=(\kb{g}{e}+\sqrt{2}\kb{e}{f}+\sqrt{3}\kb{f}{h}+\dots)\otimes\sum_{n}\kb{n}{n}.\label{eq:b}
\end{align} 
Consequently, the two-tone drive Hamiltonian in the interaction picture can be made time-independent by neglecting non-resonant terms. By also assuming $\overline{\omega}_{{\text{f}},0}-\overline{\omega}_{{\text{e}},0}\approx\overline{\omega}_{{\text{f}},1}-\overline{\omega}_{{\text{e}},1}$ and $\overline{\omega}_{{\text{f}},0}-\overline{\omega}_{{\text{g}},1}\approx\overline{\omega}_{{\text{f}},1}-\overline{\omega}_{{\text{g}},2}$, up to the three-excitation subspace, it reads
\begin{align}
    \hH/\hbar =&\ -\text{i}\frac{\Omega_{\rm{ef}}}{2}\left(\kb{\overline{e,0}}{\overline{f,0}}+\kb{\overline{e,1}}{\overline{f,1}}\right)-\text{i}\frac{\Omega_{\rm{f0g1}}}{2}\left(\kb{\overline{g,1}}{\overline{f,0}}+\sqrt{2}\kb{\overline{g,2}}{\overline{f,1}}\right) +\ \text{H.c.},\label{eq:H}
\end{align}
where $\Omega_{\rm{ef}}$ and $\Omega_{\rm{f0g1}}$ are the corresponding Rabi angular frequencies of the drives. To account for shifts of the energy levels arising from the approximations in Eq.~\eqref{eq:H} and from different physical phenomena, such as drive-induced and bath-induced shifts, we add the small correction term to the energy levels, $\hH_{\delta}=\hbar\sum_{j,n}\overline{\delta}_{{j},n}\ket{\overline{j,n}}\bra{\overline{j,n}}$. Notably, the incorporation of frequency shifts to the driven system Hamiltonian is important for accurately estimating the stabilization times and the steady state of quantum systems~\cite{Teixeira2022}.

Next, we focus on the thermal emission, absorption, and pure dephasing terms in Eq.~\eqref{eq:LME0}. Up to the three-excitation subspace, these processes are, respectively, expressed as
\begin{align}
    \mathcal{L}_{\downarrow}(\hrh) =&\  \Gamma_{\rm{eg}}\mathcal{D}\left(\kb{\overline{g,0}}{\overline{e,0}}+\kb{\overline{g,1}}{\overline{e,1}}+\kb{\overline{g,2}}{\overline{e,2}}\right)\hrh +\Gamma_{\rm{fe}}\mathcal{D}\left(\kb{\overline{e,0}}{\overline{f,0}}+\kb{\overline{e,1}}{\overline{f,1}}\right)\hrh+\Gamma_{\rm{hf}}\mathcal{D}\left(\kb{\overline{f,0}}{\overline{h,0}}\right)\hrh\nonumber\\
    {} &+\kappa_{\downarrow}\mathcal{D}\left(\kb{\overline{g,0}}{\overline{g,1}}+\kb{\overline{e,0}}{\overline{e,1}}+\kb{\overline{f,0}}{\overline{f,1}}+\sqrt{2}\kb{\overline{g,1}}{\overline{g,2}}+\sqrt{2}\kb{\overline{e,1}}{\overline{e,2}}+\sqrt{3}\kb{\overline{g,2}}{\overline{g,3}}\right)\hrh,\nonumber\\
     \mathcal{L}_{\uparrow}(\hrh) =&\  \Gamma_{\rm{ge}}\mathcal{D}\left(\kb{\overline{e,0}}{\overline{g,0}}+\kb{\overline{e,1}}{\overline{g,1}}+\kb{\overline{e,2}}{\overline{g,2}}\right)\hrh +\Gamma_{\rm{ef}}\mathcal{D}\left(\kb{\overline{f,0}}{\overline{e,0}}+\kb{\overline{f,1}}{\overline{e,1}}\right)\hrh+\Gamma_{\rm{fh}}\mathcal{D}\left(\kb{\overline{h,0}}{\overline{f,0}}\right)\hrh\nonumber\\
    {} &+\kappa_{\uparrow}\mathcal{D}\left(\kb{\overline{g,1}}{\overline{g,0}}+\kb{\overline{e,1}}{\overline{e,0}}+\kb{\overline{f,1}}{\overline{f,0}}+\sqrt{2}\kb{\overline{g,2}}{\overline{g,1}}+\sqrt{2}\kb{\overline{e,2}}{\overline{e,1}}+\sqrt{3}\kb{\overline{g,3}}{\overline{g,2}}\right)\hrh,\nonumber \\
    \mathcal{L}_{\phi}(\hrh) = &\ \gamma_{\rm{eg}}^{\phi}\mathcal{D}\left(\kb{\overline{e,0}}{\overline{e,0}}+\kb{\overline{e,1}}{\overline{e,1}}+\kb{\overline{e,2}}{\overline{e,2}}-\kb{\overline{g,0}}{\overline{g,0}}-\kb{\overline{g,1}}{\overline{g,1}}-\kb{\overline{g,2}}{\overline{g,2}}-\kb{\overline{g,3}}{\overline{g,3}}\right)\hrh/2\nonumber\\
    {}&+\gamma_{\rm{fe}}^{\phi}\left(\kb{\overline{f,0}}{\overline{f,0}}+\kb{\overline{f,1}}{\overline{f,1}}-\kb{\overline{e,0}}{\overline{e,0}}-\kb{\overline{e,1}}{\overline{e,1}}-\kb{\overline{e,2}}{\overline{e,2}}\right)\hrh/2\nonumber\\
    {}&+\gamma_{\rm{hf}}^{\phi}\left(\kb{\overline{h,0}}{\overline{h,0}}-\kb{\overline{f,0}}{\overline{f,0}}-\kb{\overline{f,1}}{\overline{f,1}}\right)\hrh/2,
    \label{eq:LME} 
\end{align}
where $\mathcal{D}(\hO)\hrh=\hO\hrh\hdgg{O}-\{\hdgg{O}\hO,\hrh\}/2$.

In Eqs.~\eqref{eq:LME}, we assume that each superoperator $\mathcal{D}$ describes the exchange of, at most, one excitation between the system and its baths. Multi-photon processes are known to take place at much lower rates than single-photon processes, and hence they are neglected. This exchange primarily occurs between the transmon-like levels $\ket{\overline{e,n}}\leftrightarrow\ket{\overline{g,n}}$, $\ket{\overline{f,n}}\leftrightarrow\ket{\overline{e,n}}$, $\ket{\overline{h,n}}\leftrightarrow\ket{\overline{f,n}}$, and the auxiliary-resonator-like levels $\ket{\overline{j,n+1}}\leftrightarrow\ket{\overline{j,n}}$. 
This assumption is motivated by the dispersive regime, which yields weakly entangled eigenstates of $\hH_0$. Consequently, we may approximate $\ket{\overline{j,n}}\approx\ket{j,n}$, with equality holding for the ground state $\ket{\overline{g,0}}$. 
Note that this factorization may only be applicable in the treatment of the open dynamics and does not hold for the Hamiltonian terms; otherwise, the $f0$--$g1$-like transitions could not be driven. 
A comparison between the master equations in the bare and dressed basis, as well as the impact of the readout resonator on reset, are left for future work.

The thermal emission and absorption rates in Eqs.~\eqref{eq:LME} can be conveniently written in terms of the decay rates $\gamma_{jk}$ and $\kappa$ as 
\begin{align}
    \Gamma_{\rm{eg}} &= \gamma_{\rm{eg}}(\bar{n}_{\rm{eg}}+1),\ \ \Gamma_{\rm{ge}} = \gamma_{\rm{eg}}\bar{n}_{\rm{eg}},\nonumber\\
    \Gamma_{\rm{fe}} &= \gamma_{\rm{fe}}(\bar{n}_{\rm{fe}}+1),\ \ \Gamma_{\rm{ef}} = \gamma_{\rm{fe}}\bar{n}_{\rm{fe}},\nonumber\\
    \Gamma_{\rm{hf}} &= \gamma_{\rm{hf}}(\bar{n}_{\rm{hf}}+1),\ \ \Gamma_{\rm{fh}} = \gamma_{\rm{hf}}\bar{n}_{\rm{hf}},\nonumber\\
    \kappa_{\downarrow} &= \kappa(\bar{n}_{\rm{r}}+1),\ \ \kappa_{\uparrow} = \kappa\bar{n}_{\rm{r}},\label{eq:rates}
\end{align}
where $\bar{n}_{jk}$ and $\bar{n}_{\rm{r}}$ are the thermal occupation numbers at the transmon and auxiliary resonator frequencies.

Using the notation of Table~\ref{tb:params}, we identify the main angular transition frequencies of the considered system as $\overline{\omega}_{{jk}}=\overline{\omega}_{{k},0}-\overline{\omega}_{{j},0}$, $\overline{\omega}_{\text{r}}=\overline{\omega}_{{g},n+1}-\overline{\omega}_{{g},n}$, and $\overline{\omega}_{\text{f0g1}}=\overline{\omega}_{\text{f},0}-\overline{\omega}_{\text{g},1}$, which are obtained through spectroscopic measurements. While the decay rate $\gamma_{\text{eg}}$ is obtained from a time-domain measurement, we use the values $\gamma_{\text{fe}}=2\gamma_{\text{eg}}$ and $\gamma_{\text{hf}}=3\gamma_{\text{eg}}$, which are expected from theory. In addition, we employ $\gamma_{jk}^{\phi}=4\gamma_{jk}$, assuming that pure dephasing is a dominant source of decoherence in the system, as verified in previous independent measurements. 

The Liouvillian model described above is utilized for fitting to the time-dependent readout signals in Figs.~\ref{fig:twotoned} and~\ref{fig:twotonef}, for the master equation simulations in Figs.~\ref{fig:sspexcg}--\ref{fig:sspexcl}, and for the Liouvillian analysis of Fig.~\ref{fig:Lpexcss}.  In these cases, the eigenspectra and the solution of Eq.~\eqref{eq:LME0} are obtained through the vectorization of the Liouvillian using the \textit{Melt} library in Mathematica~\cite{Melt}.

Specifically in Fig.~\ref{fig:twotoned}, we set the initial state to $\hrh(0)=\ket{\overline{f,0}}\bra{\overline{f,0}}$ with $\Omega_{\rm{ef}}=0$, and use
Eq.~\eqref{eq:LME0} truncated at the fourth level. To extract $\Omega_{\rm{f0g1}}$ for different voltages $V_{\rm{f0g1}}$ from the readout signal of the transmon state $\ket{f}$, we use the fitting function $s_{\text{f}}(t)=a_{\text{f}}p_{\text{f}}^{\text{theory}}(t)+ b_{\text{f}}$, where $a_{\text{f}}$ and $b_{\text{f}}$ are real-valued free parameters, and $p_{\text{f}}^{\text{theory}}(t)$ is defined in Eq.~\eqref{eq:pj_th}. From this fitting, we also obtain the auxiliary-resonator decay rates $\kappa$, the averaged value of which is shown in Table~\ref{tb:params}. We proceed in a similar way to extract the dependence of $\Omega_{\rm{ef}}$ on $V_{\rm{ef}}$ in Fig.~\ref{fig:twotonef}, by setting $\hrh(0)=\ket{\overline{e,0}}\bra{\overline{e,0}}$, $\Omega_{\rm{f0g1}}=0$, and using the fitting function $s_{\text{e}}(t)=a_{\text{e}}p_{\text{e}}^{\text{theory}}(t)+ b_{\text{e}}$, with new real-valued free parameters $a_{{e}}$ and $b_{{e}}$. 

Importantly, although the averaged signal amplitudes do not provide direct information about the transmon populations, they allow for an unbiased and fast estimate of the Rabi frequencies, helping to identify optimized reset scenarios in a more efficient way. We verified that the averaged signal amplitudes in the corresponding calibration measurements followed the above-mentioned trend $s_{i}(t)=a_{i}p_{i}(t)+b_{i}$, $i=\rm{e},\rm{f}$. The parameters $a_i$ and $b_i$ take into account for SPAM errors, thus also capturing the contributions of the other transmon eigenstates.

For the master equation simulations, we truncate the model at the tenth level and set the initial states $\hrh_{\text{g}}(0)=\hrh_{\text{thermal}}$ in Figs.~\ref{fig:sspexcg} and~\ref{fig:sspexch}, $\hrh_{\text{e}}(0)=\hsgx^{\text{ge}} \hrh_{\text{thermal}}\hsgx^{\text{ge}}$ in Figs.~\ref{fig:sspexci} and~\ref{fig:sspexcj}, and $\hrh_{\text{f}}(0)=\hsgx^{\text{ef}}\hsgx^{\text{ge}} \hrh_{\text{thermal}}\hsgx^{\text{ge}}\hsgx^{\text{ef}}$ in Figs.~\ref{fig:sspexck} and~\ref{fig:sspexcl}, where $\hsgx^{jk}$ is the Pauli $\hsgx$ operator in the corresponding transmon subspace and we have assumed a thermal natural initial state of the form $\hrh_{\text{thermal}}=\text{e}^{-\hH_0/(k_{\text{B}}T)}/Z$. This state exhibits overlap fidelity of $>99.99\%$ with the steady state of Eq.~\eqref{eq:LME0} without two-tone driving. The remaining parameters in these simulations are given as in Tables~\ref{tb:params} and~\ref{tb:configs}. To better match the experimental data, we have also included additional frequency corrections to the $\ket{\overline{e,0}}$, $\ket{\overline{g,1}}$, and $\ket{\overline{f,0}}$ levels in the simulations with two-tone driving, obtaining the optimized values $\overline{\delta}_{\text{e},0}=-2.0 $~MHz and $\overline{\delta}_{\text{g},1}=0.8$~MHz for configurations \textbf{B} and \textbf{D}, $\overline{\delta}_{\text{f},0}=-1.2$~MHz for configuration \textbf{B}, and $\overline{\delta}_{\text{f},0}=-1.6$~MHz for configuration \textbf{D}.

\subsection*{System readout}

The system readout is mostly carried out through standard dispersive techniques~\cite{Blais2021}. A $2$-{\textmu}s pulse is applied to the input port of the readout line, which is coupled to the readout resonator. The signal transmitted through the readout line is digitized after an amplification chain composed of a traveling-wave parametric amplifier (TWPA) at the millikelvin stage, and low-noise amplifiers at $50$~K and at room temperature. Ensemble-averaged measurements were first carried out with readout frequency $f_{\text{RO}}=7.4372$~GHz, which was later optimized to $f_{\text{RO}}=7.4376$~GHz for single-shot measurements. Such a frequency provided the best separation in the IQ plane for time traces integrated from $t_0=0.4$~{\textmu}s to $t_{\rm{RO}}=1.0$~{\textmu}s of the readout pulse. 

For an ideal preparation of the $\ket{e}$ state, for example, the decay-induced readout error is $1-\exp(-\gamma_{\rm{eg}}t_0)\approx 6\%$ at the beginning of the pulse and $1-\exp(-\gamma_{\rm{eg}}t_{\rm{RO}})\approx 14\%$ at the end of the pulse. Since the readout time is fixed, its error presents as a constant factor to the measured population. Therefore, it does not affect the dynamics, including the time constants and decay rates obtained from Figs.~\ref{fig:qcr}--\ref{fig:sspexc}.

In all single-shot measurements, we assume that the point clouds are well-represented by a weighted sum of four two-dimensional (2D) Gaussian distributions, 
\begin{align}
    G(\mathbf{x}) = \sum_{{j}=\text{g},\text{e},\text{f},\text{h}}
     \frac{w_{{j}}}{2\pi|\Sigma_{{j}}|^{1/2}} \exp\left[-\frac{1}{2} (\mathbf{x} - \mathbf{\mu}_{j})^{\top} \Sigma_{{j}}^{-1} (\mathbf{x} - \mathbf{\mu}_{j})\right],\label{eq:Gaussians}
\end{align}
where $\mathbf{x}$ is a 2D vector describing the coordinates of the integrated points in the IQ plane. Each Gaussian component is characterized by a mean vector $\mathbf{\mu}_{j}$, covariance matrix $\Sigma_{{j}}$, and weight $w_{{j}}$, which depend on the physical properties of the device, as well as the readout probe frequency and integration time.

We use the \textit{scikit-learn}~\cite{scikit-learn} library in Python to estimate the parameters of $G(\mathbf{x})$ based on the point distribution at $t_{\text{r}}=0$. The ellipses corresponding to the four-lowest levels of the transmon are subsequently obtained, as shown in Figs.~\ref{fig:sspexca}--\ref{fig:sspexcf}. For each set of driving configurations of Table~\ref{tb:configs} and reset lengths $t_{\text{r}}$, the populations of the transmon states are then estimated with the expression 
\begin{align}
    p_{j} = \frac{N_{{j}}}{N_{\text{g}}+N_{\text{e}}+N_{\text{f}}+N_{\text{h}}},\label{eq:pj_exp} 
\end{align}
where $N_{{j}}$ is the number of points inside the $1\sigma$-boundary ellipse corresponding to the state $\ket{j}$. We observe that the $\ket{h}$-ellipse is the largest, possibly due to inaccuracies resulting from the few number of points and the existence of the high-lying transmon states that were not considered. Nonetheless, the ratio of points inside and outside the ellipses remains close to $39.35\%$, characteristic from a 2D Gaussian distribution~\cite{Bajorski2012}. Note that this method does not take into account all state-assignment errors, and hence further single-shot-readout optimization may be achieved through more advanced machine learning algorithms~\cite{Lienhard2022,Navarathna2021,Chen2023}.

\bibliography{refs}



\section*{Acknowledgements}


This work was funded by the Academy of Finland Centre of Excellence program (project Nos. 352925, and
336810) and grant Nos. 316619 and 349594 (THEPOW).
We also acknowledge funding from the European Research Council under Advanced Grant No. 101053801
(ConceptQ). We thank Matti Silveri, Joachim Ankerhold, Gianluigi Catelani, Hao Hsu, Tapio Ala-Nissila, Shuji Nakamura for scientific discussions, and Leif Grönberg for Nb deposition.

\section*{Author contributions statement}


W.T. characterized the sample together with T.M., A.V., H.K., and M.T., conducted the QCR-assisted two-tone experiment, and carried out the data analysis. T.M. designed and carried out the electromagnetic simulations of the device, fabricated the sample, and built the experimental setup. A.G. contributed to the measurement code for single-shot readout. S.K. and A.S. calibrated the TWPA pump. V.V. contributed to the theory and design of the single-junction QCR. M.M. conceived the experiment and supervised the project. The manuscript was written by W.T., T.M., and M.M., with comments from all authors.

\section*{Additional information}

\textbf{Competing interests} The authors declare no competing interests. 

\noindent\textbf{Data availability} The datasets generated during and/or analysed during the current study are available from the corresponding author on reasonable request.

\end{document}